\documentclass[10pt,conference,compsocconf]{IEEEtran}  
\IEEEoverridecommandlockouts

\usepackage{xspace}
\usepackage{amsmath, amssymb}
\usepackage{mathptmx}
\usepackage{nicefrac}
\usepackage{booktabs}
\usepackage{psfrag}
\usepackage{listings}
\usepackage{tikz,pgfplots}
\usetikzlibrary{pgfplots.groupplots}
\usepackage{url}

\usepackage{enumitem}

\usepackage[nolist,nohyperlinks]{acronym}
\acrodef{WSN}[WSN]{Wirless Sensor Network}

\usepackage{tikz,xstring,siunitx,pgfplots,psfrag}
\usepackage{pgfplotstable}
\usepackage{ifthen}

\usetikzlibrary{shapes,arrows}
\usetikzlibrary{calc,patterns,decorations.pathmorphing,decorations.markings}
\usetikzlibrary{positioning,automata}
\usetikzlibrary{pgfplots.groupplots}

\tikzstyle{block} = [draw, rectangle, minimum height=1em, minimum width=1em]
\tikzstyle{sum} = [draw, circle, node distance=1.5cm]
\tikzstyle{input} = [coordinate]
\tikzstyle{output} = [coordinate]

\usepackage{color}

\definecolor{mygreen}{rgb}{0,0.6,0}
\definecolor{mygray}{rgb}{0.5,0.5,0.5}
\definecolor{mymauve}{rgb}{0.58,0,0.82}

\usepackage{listings}
\usepackage[colorinlistoftodos, textwidth=20mm, textsize=footnotesize]{todonotes}

\begin{document}
\onecolumn

\section*{IEEE Copyright Notice}
Copyright (c) 2015 IEEE.
Personal use of this material is permitted. Permission from IEEE must be
obtained for all other uses, in any current or future media, including
reprinting/republishing this material for advertising or promotional purposes,
creating new collective works, for resale or redistribution to servers or lists,
or reuse of any copyrighted component of this work in other works.\\
\\
F. Terraneo, A. Leva, S. Seva, M. Maggio, A. V. Papadopoulos, "Reverse Flooding: exploiting radio interference for efficient propagation delay compensation in WSN clock synchronization" IEEE Real-Time Systems Symposium (RTSS), San Antonio, Texas, December 2015.\\
\\
https://doi.org/10.1109/RTSS.2015.24
\twocolumn
\clearpage

\title{\LARGE \bf Reverse Flooding: exploiting radio interference for efficient propagation delay compensation in WSN clock synchronization
\thanks{This work was partially supported by the LCCC Linnaeus 
  and ELLIIT Excellence Centers and the Swedish Research Council
  (VR) for the projects ``Cloud Control'' and ``Power and temperature control 
  for large-scale computing infrastructures''.}}

\author{\IEEEauthorblockN{Federico Terraneo\IEEEauthorrefmark{1}, Alberto Leva\IEEEauthorrefmark{1}, Silvano Seva\IEEEauthorrefmark{1}, Martina Maggio\IEEEauthorrefmark{2}, Alessandro Vittorio Papadopoulos\IEEEauthorrefmark{2}}
  \IEEEauthorblockA{\IEEEauthorrefmark{1}Politecnico di Milano, Milano, Italy}
  \IEEEauthorblockA{\IEEEauthorrefmark{2}Lund University, Lund, Sweden}}

\maketitle
\thispagestyle{empty}
\pagestyle{empty}

\begin{abstract}
Clock synchronization is a necessary component in modern distributed
systems, especially \acp{WSN}. Despite the great effort and the
numerous improvements, the existing synchronization schemes do not yet
address the cancellation of propagation delays. Up to a few years ago,
this was not perceived as a problem, because the time-stamping
precision was a more limiting factor for the accuracy achievable with
a synchronization scheme.
However, the recent introduction of efficient flooding schemes based on
constructive interference has greatly improved the achievable
accuracy, to the point where propagation delays can effectively become
the main source of error.

In this paper, we propose a method to estimate and compensate for the
network propagation delays. Our proposal does not require to maintain
a spanning tree of the network, and exploits constructive interference
even to transmit packets whose content are slightly different. To show
the validity of the approach, we implemented the propagation delay
estimator on top of the FLOPSYNC-2 synchronization scheme.
Experimental results prove the feasibility of measuring propagation
delays using off-the-shelf microcontrollers and radio transceivers,
and show how the proposed solution allows to achieve sub-microsecond
clock synchronization even for networks where propagation delays are
significant.

\end{abstract}

\acresetall

\section{Introduction}
\label{sec:Intro}

Clock synchronization in distributed systems is a problem with a long
history~\cite{bib:KopetzEtAl-1987a, bib:Lamport-1978a}. Recently, the
diffusion of \acp{WSN} has drawn more attention to specific variants
of the problem. Indeed, there are many different clock synchronization
problems, depending on the settings and the desired properties that a
synchronization scheme should achieve. The following work is cast in
the framework of \emph{multi-hop master-slave} clock synchronization.

In the specific problem addressed in this paper, the network is
composed by a certain number of slave nodes and a single master
node. The slave nodes are organized in hops, around the master node,
which is the only node belonging to Hop 0. The master node has a
limited transmission range and directly reaches nodes belonging to Hop
1.  In turn, nodes belonging to Hop 1 can re-broadcast the messages
received from the master node and reach nodes that are farther away,
not belonging to the master range. These nodes belong to Hop 2 and can
receive master communications only due to the re-transmission of the
nodes that directly receive the master node packets. The procedure
can be repeated adding an arbitrarily large number of nodes and
hops. The problem, in this case, is to
synchronize the clocks of all the slave nodes to the clock of the
master node, irregardless of the hop they belong to.

While there are quite a few solutions for the master-slave multi-hop
synchronization problem~\cite{bib:MarotiEtAl-2004a, bib:ElsonEtAl-2002a, bib:PingSu-2003a, bib:TerraneoEtAl-2014a}, none
of them takes directly into account the propagation delay of the
packets on the network. Historically, the propagation delay was not a
big issue. The most limiting factor to what a synchronization scheme
could achieve was the time-stamping precision. However, quoting
from~\cite{bib:SchmidEtAl-2010b}, ``\emph{While [propagation delay] is
  not a problem if time-stamping precision is worse than about 1
  $\mu$s, it starts to be a significant source of error at appreciably
  finer precisions}''.

The recent introduction of flooding schemes based on constructive
interference like Glossy~\cite{bib:FerrariEtAl-2011a}, has changed the
\ac{WSN} synchronization \emph{scenario}. Making the flooding
mechanism insensitive to MAC-induced delays has in fact allowed to
disseminate timing information with unprecedented precision.
Moreover, skew compensation techniques have also advanced, allowing
for sub-microsecond precision also in ultra-low
power networks~\cite{bib:TerraneoEtAl-2014a}. In these conditions,
propagation delays become a problem to be addressed.

In this paper we present a delay compensation method that can be built
on top of any asymmetric master-slave clock synchronization scheme
based on constructive interference flooding. This paper makes the
following contributions.
\begin{itemize}[noitemsep,topsep=0pt]
\item It enhances a master-slave synchronization scheme, exploiting the
  proposed method to estimate and compensate for the propagation delay
  from the master node to any slave node in a \ac{WSN} \emph{without
    the need for a spanning tree} of the network.
\item As a second methodological contribution, it presents a technique
  to allow the concurrent transmission of multiple packets having
  \emph{a different payload}. The method is capable of exploiting the
  constructive interference despite the different payloads, resulting
  in an intelligible message for the intended recipient.
\end{itemize}
To show the applicability of the technique we implemented the proposed
delay compensator on top of the FLOPSYNC-2 synchronization
scheme~\cite{bib:TerraneoEtAl-2014a}, showing how the method improves
time synchronization in a \ac{WSN} where sensors are deployed within a
radius that causes propagation delay to be the major source of error.

\section{Problem statement}
\label{sec:PSanddBD}

Consider a multi-hop \ac{WSN} with one master node and a certain
number of slave nodes. Each node is equipped with a synchronization
scheme based on a MAC-level flooding scheme like
Glossy~\cite{bib:FerrariEtAl-2011a}.

If we assume that nodes do not move, it is possible to define a
\emph{flooding graph} for the network. The flooding graph is a
subgraph of the directed graph connecting all nodes of hop $h$ with
all nodes of hop $h+1$ for each network hop. The edges missing in the
flooding graph with respect to the complete one are due to
distances. In fact, if a receiver node is not in the radio range of
the transmitter one, the corresponding edge is removed from the
flooding graph.

\begin{figure}[t]
 \begin{center}
  \usetikzlibrary{matrix}
\usetikzlibrary{arrows}
\usetikzlibrary{calc}

\tikzstyle{nnode}   = [draw,circle,thick,inner sep=0mm,minimum size=3.2mm]
\tikzstyle{alink}   = [->] % active link
\tikzstyle{ilink}   = [->,dashed] % inactive link

% ARC BY CENTRER, RADIUS AND ANGLES
% {draw options}{xcen}{ycen}{radius}{start angle}{end angle}
\def\arc_cra#1#2#3#4#5#6 {
  \draw[#1] ({#2+#4*cos(#5)},{#3+#4*sin(#5)}) arc (#5:#6:#4);
}

% NODE WITH CIRCULAR RADIO RANGE
% {name}{caption}{x}{y}{node draw options}{range radius}{circle draw options}{circle fill options}
\def\nnode_rrange#1#2#3#4#5#6#7#8 {
  \draw[#8] (#3,#4) circle (#6);
  \draw[#7] (#3,#4) circle (#6);
  \node[nnode,#5] at (#3,#4) (#1) {{\tiny #2}};
}

\begin{tikzpicture}[scale=1.00]

\node[black] (h0) at (01.85,08.90) {{\small Hop 0}};

% master
\arc_cra{densely dotted,color=black} {04.50}{08.10}{2.25}{-200}{35}
\node [nnode,color=black] at (04.50,08.10) (M) {{\tiny M}};

\node[black!40!blue] (h1) at (03.50,09.25) {{\small Hop 1}};

% node 1
\nnode_rrange{N1}{1}{03.55}{06.45}{color=black!40!blue}{2.25}
   {color=blue,dotted}{fill=blue,opacity=0.05}
% node 2
\nnode_rrange{N2}{2}{04.20}{06.45}{color=black!40!blue}{2.25}
   {color=blue,dotted}{fill=blue,opacity=0.05}

% node 3
\nnode_rrange{N3}{3}{05.00}{07.15}{color=black!40!blue}{2.25}
   {color=blue,dotted}{fill=blue,opacity=0.05}

\node[black!20!red] (h2) at (08.40,09.15) {{\small Hop 2}};

%node 4
\nnode_rrange{N4}{4}{04.00}{05.55}{color=black!20!red}{2.25}
   {color=red,dotted}{fill=red,opacity=0.05}

%node 5
\nnode_rrange{N5}{5}{05.95}{05.60}{color=black!20!red}{2.25}
   {color=red,dotted}{fill=red,opacity=0.05}

%node 6
\nnode_rrange{N6}{6}{06.90}{07.00}{color=black!20!red}{2.25}
   {color=red,dotted}{fill=red,opacity=0.05}

%node 7
\node[nnode,color=black!50!green] at (03.70,03.70) (N7) {{\tiny 7}};

%node 8
\node[nnode,color=black!50!green] at (07.50,05.50) (N8) {{\tiny 8}};

%node 9
\node[nnode,color=black!50!green] at (08.00,08.00) (N9) {{\tiny 9}};

\draw[alink] (M.230) -- (N1.60);
\draw[alink] (M.260) -- (N2.80);
\draw[alink] (M.300) -- (N3.120);

\draw[alink,color=black!40!blue] (N1.290) -- (N4.120);
\draw[alink,color=black!40!blue] (N2.260) -- (N4.80);
\draw[alink,color=black!40!blue] (N2.-20) -- (N5.140);
\draw[alink,color=black!40!blue] (N3.-60) -- (N5.110);
\draw[alink,color=black!40!blue] (N3.-10) -- (N6.170);

\draw[alink,color=black!20!red] (N4.260) -- (N7.85);
\draw[alink,color=black!20!red] (N5.-10) -- (N8.175);
\draw[alink,color=black!20!red] (N6.-60) -- (N8.115);
\draw[alink,color=black!20!red] (N6.40) -- (N9.220);

\end{tikzpicture}
 \end{center}
 \caption{Flooding graph example with nodes' radio ranges.}
 \label{fig:PSanddBD-FGex}
\end{figure}
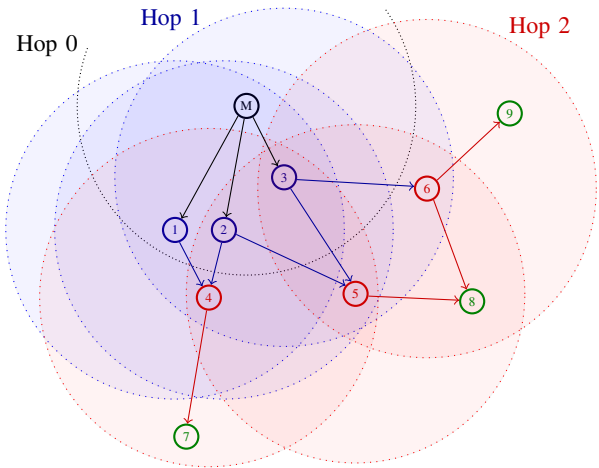

Figure~\ref{fig:PSanddBD-FGex} shows an example. The Hop 0 is composed
by the master node, marked with the letter M. Nodes 1, 2 and 3 belongs
to Hop 1, since they are in the radio range of the master node
(depicted in the figure with a dotted circle centered on the master
node). Nodes 4, 5 and 6 belong to Hop 2, since they are in the radio
range of at least one of the nodes belonging to Hop 1. The remaining
nodes, 7 and 8, belong to Hop 3 -- their radio range is not shown to
simplify the picture. Hop 3 illustrates also another characteristic of
the network. Nodes belonging to a hop do not, in general, receive packets
from all the nodes in the previous one. In the case of Hop 3, node 7
belongs to the radio range of node 4 (and would not receive any packet
from node 5 and 6), node 8 belongs to the radio range of both node 5
and node 6 (but not of node 4) and node 9 receives packets only from
node 6. As nodes belonging to a hop can receive packets from
multiple nodes belonging to the previous one, in general the
flooding graph is not a tree. This motivates the need to exploit the constructive
interference between packets transmitted by nodes that are close to
each other.

We would like each node to be able to reliably estimate the
\emph{propagation delay} from the master node to it. This is
complicated by the fact that the flooding graph is not a tree and that
nodes do not possess knowledge about its structure. In fact, with a
spanning tree and knowledge about the tree structure, each node $i$
could estimate via round-trip delay measurements the propagation delay
$\delta_{i \rightarrow p}$ from its \emph{sole} predecessor $p$ and could
simply ask to the predecessor the cumulated delay $\delta_{M
  \rightarrow p}$ from the master $M$ to $p$. Then the node $i$ would
have an estimate of the delay from the master: $\delta_{M \rightarrow
  i} = \delta_{i \rightarrow p} + \delta_{M \rightarrow p}$. We
propose a solution based on the same principles, that takes into
account the nature of the graph and the constructive interference in
the transmissions.

The main difficulties when dealing with the flooding graph and
constructive interference are the following.
\begin{enumerate}
\item The node $i$ does not have a single predecessor $p$ but a set of
  predecessors $\mathcal{P}_i = \{ p_1, \dots, p_n \}$ and receives
  the flooded timing information by the entire set of predecessors; round-trip
  measurements in such a \emph{scenario} need to take this into account.
\item When node $i$ queries nodes in $\mathcal{P}_i$ for their cumulated
  delay from the master node, they will simultaneously send back
  (potentially) different responses, that should be fused in a
  meaningful manner.
\item The node $i$ does not know which are the nodes that form the set
  $\mathcal{P}_i$. More in general nodes should query one another for
  round-trip times along the flooding graph, but none of them knows
  the structure of the graph. One of the major strength of
  interference-based flooding -- not knowing which are the nodes that
  constructively interfere to provide timing information -- turns here
  into a problem.
\end{enumerate}

In Section~\ref{sec:RoundTrip} we describe how a node can estimate the
propagation delay from its predecessor set $\mathcal{P}_i$ without
knowing which nodes belong to it. In
Section~\ref{sec:BarGraphEncoding} we propose an encoding method to
fuse the different responses about the cumulated delay from the master
node. Neither of these two require knowledge of the flooding graph.
Section~\ref{sec:FullScheme} describes a suitable Time Division
Multiple Access (TDMA) method so that all the nodes of the
network can perform the two tasks just mentioned.
Section~\ref{sec:ExpResults} presents and discusses experimental results.

\section{Last-hop delay measurement}
\label{sec:RoundTrip}

In this section we show how a node $i$ can measure the propagation
delay from its predecessor set $\mathcal{P}_i$ (in contrast to the
propagation delay from a single node). The first step is
to define which nodes belong to the predecessor set
$\mathcal{P}_i$. In so doing, we ensure that the node does not need
any information about the flooding graph.

\begin{figure}[t]
 \begin{center}
  \usetikzlibrary{matrix}
\usetikzlibrary{arrows}
\usetikzlibrary{calc}

\tikzstyle{nnode}   = [draw,circle,thick,inner sep=0mm,minimum size=3.2mm]
\tikzstyle{alink}   = [->] % active link
\tikzstyle{ilink}   = [->,dashed] % inctive link

% ARC BY CENTRER, RADIUS AND ANGLES
% {draw options}{xcen}{ycen}{radius}{start angle}{end angle}
\def\arc_cra#1#2#3#4#5#6 {
  \draw[#1] ({#2+#4*cos(#5)},{#3+#4*sin(#5)}) arc (#5:#6:#4);
}

% NODE WITH CIRCULAR RADIO RANGE
% {name}{caption}{x}{y}{node draw options}{range radius}{circle draw options}{circle fill options}
\def\nnode_rrange#1#2#3#4#5#6#7#8 {
  \draw[#8] (#3,#4) circle (#6);
  \draw[#7] (#3,#4) circle (#6);
  \node[nnode,#5] at (#3,#4) (#1) {{\tiny #2}};
}

\begin{tikzpicture}[scale=1.00]

% master
\arc_cra{densely dotted,color=black} {04.50}{08.10}{2.25}{-190}{30}
\node [nnode,color=black] at (04.50,08.10) (M) {{\tiny M}};

% node 1
\nnode_rrange{N1}{1}{03.55}{06.45}{color=black!40!blue}{2.25}
   {color=blue,dotted}{fill=blue,opacity=0.05}
% node 2
\nnode_rrange{N2}{2}{04.20}{06.45}{color=black!40!blue}{2.25}
   {color=blue,dotted}{fill=blue,opacity=0.05}

% node 3
\nnode_rrange{N3}{3}{05.00}{07.15}{color=black!40!blue}{2.25}
   {color=blue,dotted}{fill=blue,opacity=0.05}

%node 4
\arc_cra{densely dotted,color=red} {04.00}{05.50}{2.25}{200}{-20}
\node [nnode,color=red] at (04.00,05.50) (N4) {{\tiny 4}};

\draw[alink] (M.230) -- (N1.60);
\draw[alink] (M.260) -- (N2.80);
\draw[alink] (M.300) -- (N3.120);

\draw[alink,color=black!40!blue] (N1.290) -- (N4.120);

\draw[alink,color=black!40!blue] (N2.260) -- (N4.80);

\draw[ilink,color=black!40!blue] (N3.240) -- (N4.50);

\end{tikzpicture}
 \end{center}
 \caption{Example of flooding and the capture effect.}
 \label{fig:HopDelay}
\end{figure}
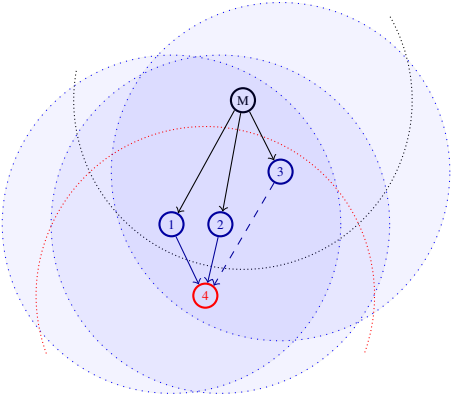

Consider node 4 in the example of Figure~\ref{fig:PSanddBD-FGex} (to
help the reader, node 4, its predecessor set and the master node are
shown in Figure~\ref{fig:HopDelay}). The node is placed in the radio
range of nodes 1, 2 and 3, but the distances between these nodes and
node 4 are different -- node 1 and 2 are closer to the node while the
distance from node 3 is larger.

During flooding, node 4 receives the packets sent concurrently by
nodes node 1 and 2 thanks to the constructive interference. However, since
node 3 is farther, its signal is received as weaker than those of
nodes 1 and 2. Due to \emph{capture
  effects}~\cite{bib:LeentvaarFlint-1976a}, the packet sent from node
3 is shadowed by the stronger signals of nodes 1 and 2. Indeed,
flooding schemes like Glossy require all nodes to transmit with the
same power~\cite{bib:FerrariEtAl-2011a}. This ensures that nodes that
are more distant, thus having a higher propagation delay, have a
weaker signal.

A node $p$ belonging to the previous hop is either at a comparable
distance with the closest nodes, having enough power to interfere
constructively, or its transmission is shadowed by capture effects. In
case the node is at a comparable distance with the closest ones, its
cumulative propagation delay from the master $\delta_{M \rightarrow
  p}$ is also comparable to the one of the closest nodes. With respect
to Figure~\ref{fig:HopDelay}, $\delta_{M \rightarrow 1}$ and
$\delta_{M \rightarrow 2}$ are comparable, while $\delta_{M
  \rightarrow 3}$ could be different, but its transmission is not
processed by node 4.
This property allows us to define the predecessor set $\mathcal{P}$ of
node $i$ as the set of the closest nodes in the previous hop that are
received with a comparable power. In the example, $\mathcal{P}_4 = \{
1, 2 \}$.

% \subsection{Round-trip delay estimation}

Once the predecessor set $\mathcal{P}_i$ is defined, node $i$ needs to
measure the last-hop propagation delay. To do so without knowing the
flooding graph, the key idea is to replicate in a round-trip
measurement the same conditions of
concurrent transmission that occur during flooding. To achieve this, we reserve a short time window after
flooding, where the MAC protocol is still disabled and the radio
channel is still clear from access contention.

\begin{figure}[t]
 \begin{center}
  \usetikzlibrary{matrix}
\usetikzlibrary{arrows}
\usetikzlibrary{calc}

\tikzstyle{nnode}   = [draw,circle,thick,inner sep=0mm,minimum size=3.2mm]
\tikzstyle{alink}   = [->] % active link
\tikzstyle{ilink}   = [->,dashed] % inctive link

% ARC BY CENTRER, RADIUS AND ANGLES
% {draw options}{xcen}{ycen}{radius}{start angle}{end angle}
\def\arc_cra#1#2#3#4#5#6 {
  \draw[#1] ({#2+#4*cos(#5)},{#3+#4*sin(#5)}) arc (#5:#6:#4);
}

% NODE WITH CIRCULAR RADIO RANGE
% {name}{caption}{x}{y}{node draw options}{range radius}{circle draw options}{circle fill options}
\def\nnode_rrange#1#2#3#4#5#6#7#8 {
  \draw[#8] (#3,#4) circle (#6);
  \draw[#7] (#3,#4) circle (#6);
  \node[nnode,#5] at (#3,#4) (#1) {{\tiny #2}};
}

\begin{tikzpicture}[scale=1.00]

% master
\node [nnode,color=black] at (04.50,08.10) (M) {{\tiny M}};

% node 1
\nnode_rrange{N1}{1}{03.55}{06.45}{color=black!40!blue}{0}
   {color=blue,dotted}{fill=blue,opacity=0.05}
% node 2
\nnode_rrange{N2}{2}{04.20}{06.45}{color=black!40!blue}{0}
   {color=blue,dotted}{fill=blue,opacity=0.05}

% node 3
\nnode_rrange{N3}{3}{05.00}{07.15}{color=black!40!blue}{0}
   {color=blue,dotted}{fill=blue,opacity=0.05}

%node 4
\nnode_rrange{N4}{4}{04.00}{05.55}{color=black!20!red}{2.25}
   {color=red,dotted}{fill=red,opacity=0.05}

%node 5
\nnode_rrange{N5}{5}{05.95}{05.60}{color=black!20!red}{0}
   {color=red,dotted}{fill=red,opacity=0.05}

%node 6
\nnode_rrange{N6}{6}{06.90}{07.00}{color=black!20!red}{0}
   {color=red,dotted}{fill=red,opacity=0.05}

%node 7
\node[nnode,color=black!50!green] at (03.70,03.70) (N7) {{\tiny 7}};

%node 8
\node[nnode,color=black!50!green] at (07.50,05.50) (N8) {{\tiny 8}};

%node 9
\node[nnode,color=black!50!green] at (08.00,08.00) (N9) {{\tiny 9}};

% {draw options}{xcen}{ycen}{radius}{start angle}{end angle}
\node [left] at (03.00,06.50) (M) {{\small Hop 1}};
\arc_cra{thick,dotted}{3.2}{8.7}{2}{-11}{-120}
\arc_cra{thick,dotted}{3.5}{8.7}{2.8}{-9}{-125}
\arc_cra{dotted}{3.5}{8.5}{4.1}{-3}{-115}

\node [right,color=red] at (06.00,04.50) {{\small Node 4 in hop 2}};
\node [right,color=red] at (05.84,04.25) {{\small sends round-trip}};
\node [right,color=red] at (05.64,04.00) {{\small request addressing}};
\node [right,color=red] at (05.40,03.75) {{\small nodes in hop 1}};

\draw[alink,thick,color=red] (N4.120) -- (N1.290);
\draw[alink,thick,color=red] (N4.80) -- (N2.260);
\draw[alink,thick,color=red] (N4.260) -- (N7.85);
\draw[alink,thick,color=red] (N4.50) -- (N3.245);
\draw[alink,thick,color=red] (N4.0) -- (N5.180);

\end{tikzpicture}
 \end{center} 
 \begin{center}
  \usetikzlibrary{matrix}
\usetikzlibrary{arrows}
\usetikzlibrary{calc}

\tikzstyle{nnode}   = [draw,circle,thick,inner sep=0mm,minimum size=3.2mm]
\tikzstyle{alink}   = [->] % active link
\tikzstyle{ilink}   = [->,dashed] % inctive link

% ARC BY CENTRER, RADIUS AND ANGLES
% {draw options}{xcen}{ycen}{radius}{start angle}{end angle}
\def\arc_cra#1#2#3#4#5#6 {
  \draw[#1] ({#2+#4*cos(#5)},{#3+#4*sin(#5)}) arc (#5:#6:#4);
}

% NODE WITH CIRCULAR RADIO RANGE
% {name}{caption}{x}{y}{node draw options}{range radius}{circle draw options}{circle fill options}
\def\nnode_rrange#1#2#3#4#5#6#7#8 {
  \draw[#8] (#3,#4) circle (#6);
  \draw[#7] (#3,#4) circle (#6);
  \node[nnode,#5] at (#3,#4) (#1) {{\tiny #2}};
}

\begin{tikzpicture}[scale=1.00]

% master
\node [nnode,color=black] at (04.50,08.10) (M) {{\tiny M}};

% node 1
\nnode_rrange{N1}{1}{03.55}{06.45}{color=black!40!blue}{2.25}
   {color=blue,dotted}{fill=blue,opacity=0.05}
% node 2
\nnode_rrange{N2}{2}{04.20}{06.45}{color=black!40!blue}{2.25}
   {color=blue,dotted}{fill=blue,opacity=0.05}

% node 3
\nnode_rrange{N3}{3}{05.00}{07.15}{color=black!40!blue}{2.25}
   {color=blue,dotted}{fill=blue,opacity=0.05}

%node 4
\nnode_rrange{N4}{4}{04.00}{05.55}{color=black!20!red}{0}
   {color=red,dotted}{fill=red,opacity=0.05}

%node 5
\nnode_rrange{N5}{5}{05.95}{05.60}{color=black!20!red}{0}
   {color=red,dotted}{fill=red,opacity=0.05}

%node 6
\nnode_rrange{N6}{6}{06.90}{07.00}{color=black!20!red}{0}
   {color=red,dotted}{fill=red,opacity=0.05}

%node 7
\node[nnode,color=black!50!green] at (03.70,03.70) (N7) {{\tiny 7}};

%node 8
\node[nnode,color=black!50!green] at (07.50,05.50) (N8) {{\tiny 8}};

%node 9
\node[nnode,color=black!50!green] at (08.00,08.00) (N9) {{\tiny 9}};

\node [right,color=darkgray] at (05.50,04.50) {{\small Nodes 5 and 7, not}};
\node [right,color=darkgray] at (05.20,04.25) {{\small in the addressed hop 1,}};
\node [right,color=darkgray] at (04.80,04.00) {{\small do not respond}};

\node [right,color=blue] at (05.70,09.35) {{\small Shadowed response}};
\node [right,color=blue] at (06.15,09.10) {{\small from node 3}};

\node [right,color=black!60!blue] at (01.46,05.50) {{\small Constructive}};
\node [right,color=black!60!blue] at (01.63,05.25) {{\small interference}};
\node [right,color=black!60!blue] at (01.82,05.00) {{\small of nodes 1}};
\node [right,color=black!60!blue] at (02.08,04.75) {{\small and 2}};

\draw[alink,thick,color=black!40!blue] (N1.290) -- (N4.120);
\draw[alink,thick,color=black!40!blue] (N2.260) -- (N4.80);
\draw[ilink,thick,color=black!40!blue] (N3.245) -- (N4.50);

\end{tikzpicture}
 \end{center}
 \caption{Round-trip measurement initiated by node 4 -- request (top)
   and response (bottom).}
 \label{fig:RoundTrip-1}
\end{figure}
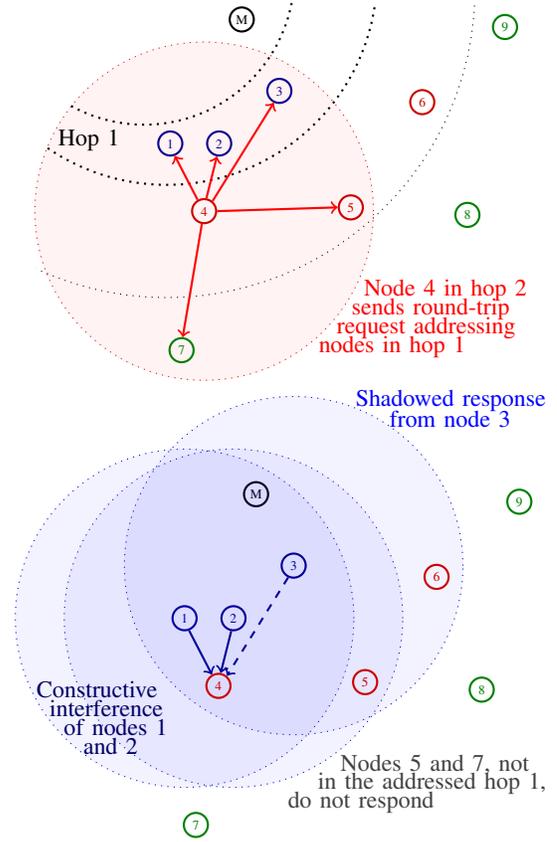

Referring again to the example in Figure~\ref{fig:HopDelay}, within
its time window, node 4 can initiate the measurement by sending a
packet with its hop number minus one (in the example, 1). The
difference with respect to standard round-trip estimation is the
packet content. While in standard round-trip estimation the
request packet contains the unique id of a node, in this case
the packet contains the hop that should respond to the message. This
is exemplified in the top part of Figure~\ref{fig:RoundTrip-1}.

Due to the radio range of the node and the definition of hops, the
packet sent by node $i$ can only be received by nodes belonging to the
previous, same and subsequent hop.
A node $m$ that receives the packet checks that the content matches its own hop
number. In case it does not, the node simply ignores the request. Otherwise,
$m$ waits for a fixed time $\tau_w$ and then replies with another
packet. As we have assumed that all nodes transmit at the same power,
the radio ranges are symmetric, and the packet sent by node 4 is
received by nodes 1, 2, 3, 5 and 7. Node 5 and 7 ignore the packet, while
the others process it. Notice that also node 3 receives the request packet.
The distance of node 3 does not make any
difference in this case, since a single packet is being sent, contrary
to multiple interfering ones. After a fixed time, nodes 1, 2 and 3
reply with an answer packet to node 4, replicating the same concurrent
transmission condition of flooding -- in this case, the response
packet sent by node 3 will again be shadowed by the stronger signal of
nodes 1 and 2 -- as shown in the bottom part of
Figure~\ref{fig:RoundTrip-1}.

The $i$-th node can measure the time difference between its packet
transmission and the reception and, knowing $\tau_w$, can estimate the
propagation delay from the nodes that belong to $\mathcal{P}_i$,
without knowing them.

\section{Cumulated delay estimation}
\label{sec:BarGraphEncoding}

Once the node $i$ has an estimate of the last-hop delay, it needs
to obtain an estimate of the sum of all the propagation delays for
each additional hop that separates it from the master. We solve the
problem recursively, querying nodes in the previous hop for their
cumulated delay from the master.

Although the capture effects ensure that constructive interference
occurs only between nodes at a comparable distance from the
receiver, one should also take into
account noise in round-trip measurements and small distance
differences. These may cause the nodes in a predecessor set to have
similar but not equal delay measurements from the master. For
example, node 1 and 2 in Figure~\ref{fig:HopDelay} may have similar
but not equal cumulated delay values, $\delta_{M \rightarrow 1} =
\delta_{M \rightarrow 2} + \varepsilon$ for small values of
$\varepsilon$. While this is not a problem for the estimate of the
last-hop delay, it becomes a problem for the cumulated delay. Since
the nodes do not have knowledge of the flooding graph, they cannot
simply query one specific node. In the example, if node 4 knew its
predecessors, it could simply ask the cumulated delay to 1 and 2
separately, and then average the response. However, node 1 and 2 will
transmit their responses concurrently.

To date, \ac{WSN} interference was studied~\cite{zhang2006infocom, liu2010icnp, boano2010wsn} and constructive interference used to
transmit the same packet. We propose a method to fuse packets with
different payloads and transmitted simultaneously from multiple
sources.

To better understand what happens when different packets are received
concurrently, it is necessary to briefly discuss the operation of an
IEEE 802.15.4 radio, which is the most common standard for WSN. A
packet is composed of a 4-byte preamble, used for the receiver to lock
on the incoming data, followed by a one-byte start frame delimiter
that marks the packet beginning. The following byte indicates the
packet length. The payload follows, and finally a two-byte Cyclic
Redundancy Check (CRC) terminates the packet. Data is grouped in 4-bit
nibbles, and for each nibble a sequence of 32 bit from a look-up table
is sent over the radio. This introduces redundancy in the transmitted
data, improving reception in adverse conditions. The receiver, at each
32 decoded bits, attempts to find which of the 16 possible sequences best
matches the received signal, and outputs the corresponding nibble.

The minimum transmission unit is one nibble, and when packets with
different payloads are sent concurrently, every nibble that has the
same value in all packets interferes constructively. On the contrary,
nibbles that have different values interfere destructively, resulting
in unpredictable nibbles being received.

\subsection{Concurrent transmission through bar graph encoding}

We propose to utilize an \emph{ad hoc} encoding, which we denote as
\emph{bar graph encoding}, that results in intelligible packets even
when some nibbles interfere destructively.

To better explain the concept behind the bar graph encoding, assume to
have an 8-bytes-long packet payload and to encode a number, bounded in
the range between 0 and 16 as the number of consecutive 0xf nibbles
starting from the beginning of the packet, leaving all other nibbles
as 0x0. The number 0 would be encoded with a packet full of 0x0, the
number 16 with a packet full of 0xf and, for example, the number 5 as
\texttt{ff$\,$ff$\,$f0$\,$00$\,$00$\,$00$\,$00$\,$00}. When sending
two different numbers, for example 5 and 8, the two packets
\begin{center}
\texttt{ff$\,$ff$\,$f0$\,$00$\,$00$\,$00$\,$00$\,$00}\\
\texttt{ff$\,$ff$\,$ff$\,$ff$\,$00$\,$00$\,$00$\,$00}
\end{center}
will be transmitted, and the generic received packet will look like
\texttt{ff$\,$ff$\,$fX$\,$XX$\,$00$\,$00$\,$00$\,$00}, with \texttt{X}
being an unpredictable value.
This encoding allows us to concurrently transmit different values,
that the radio channel itself merges. In principle, the received
unpredictable values could differ from 0x0 and 0xf, thus simplifying
the estimate of the maximum and minimum index in the packet of the
values sent concurrently. However, experimental results have shown
that with high likelihood the unpredictable nibbles are a random
pattern of either 0x0 or 0xf, while different values occur with a
significantly lower probability.

In our proposal, the nodes send the cumulated delay using the bar
graph encoding. Since the differences between the values sent by
different nodes are small, we also assume that any possible value
between the bounds of the sent one is acceptable. In the example
above, the packet decoding would be successful if any value between 5
and 8 was returned.

With bar graph encoding, packets need to be sent without a CRC,
otherwise the failed CRC due to nibble errors would result in the
packet being discarded. Although the 802.15.4 standard prescribes a
CRC at the end of each packet, radio transceivers such as the CC2520
have an option to disable its transmission.
Packet decoding and validation is implemented by identifying the
boundary from the left where two consecutive nibbles are different
from 0xf, and the boundary from the right where two consecutive
nibbles are different from 0x0. If the difference between the two
boundaries is greater than a threshold, the packet is considered
invalid and discarded. Otherwise, the average value between the two
boundaries is considered as the transmitted number from the
predecessor set. Notice that this decoding algorithm is robust to
non-consecutive nibble errors anywhere within the payload. This favors
correct reception also in adverse conditions, such as concurrent
transmission of long packets. The full C++ implementation of the
decoding algorithm used in the experimental evaluation is shown in
Listing~\ref{alg:barDecoder}.

\begin{figure}[tb]
  \lstinputlisting[language=C++,%
  caption={Bar graph encoding decoder algorithm.},%
  label={alg:barDecoder}]%
  {./sections/bar_encoding_decoder.cpp}
\end{figure}

802.15.4 packets have a maximum length of 127 bytes, thus the proposed
technique permits the transmission of a number in the range $[0,
254]$. Assuming that the timestamping resolution is $42\,$ns, as in
the experimental results section, this allows us to handle propagation
delays of up to $10.58\,\mu$s. In turn, we can synchronize nodes in a
range of about $3\,$km from the master with the maximum precision allowed by our timestamping resolution. To
handle larger distances, it is possible to lower the propagation delay
resolution. As an extreme example, with a resolution of $1\,\mu$s, the
range would extend to about $75\,$km.

The request sent by a node $i$ (4 in the example) to its predecessor
set $\mathcal{P}_i$ (in the example $\mathcal{P}_4 = \{1,2\}$) for the
cumulated delay from the master, can be made implicit in the
round-trip request packet described in
Section~\ref{sec:RoundTrip}. The answer can be piggybacked to the
round-trip answer packet, resulting in both round-trip estimation and
cumulated delay communication with a single packet exchange.

\section{The complete scheme}
\label{sec:FullScheme}

This section describes how the last-hop delay measurement and the
cumulated delay reception are composed to properly estimate the delay
from the master. We assume for the moment that a network of $n$ nodes is already
formed, and that each node has a unique id and knows its hop number, which is true if a
flooding scheme like Glossy~\cite{bib:FerrariEtAl-2011a} is used.

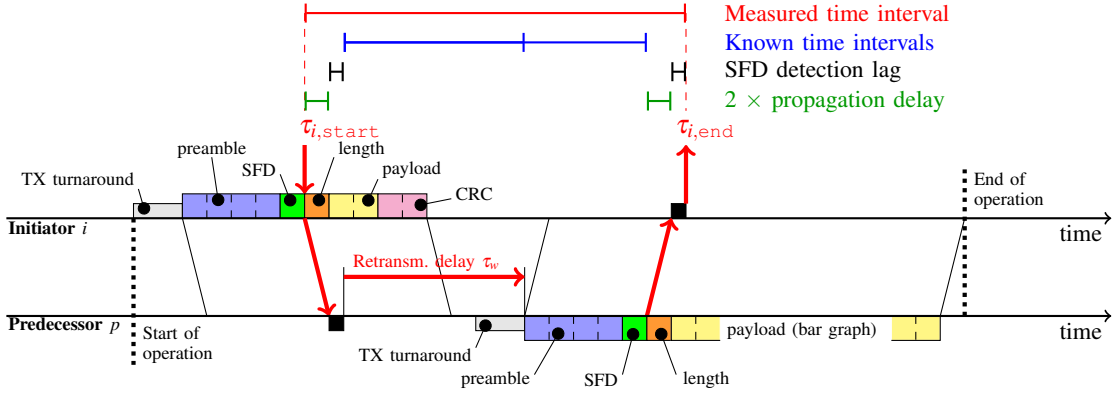
\begin{figure*}[t]
 \begin{center}
  \usetikzlibrary{matrix}
\usetikzlibrary{arrows}
\usetikzlibrary{calc}

\begin{tikzpicture}[scale=0.65]

% SLAVE ---------------------------------------------------------------------------------------------------------------

% TX turnaround
\draw[fill=gray!20] (01.00,10.00) rectangle (02.00,10.30);
\draw[*-] (01.25,10.08) -- (01.10,10.60);
\node[left] at (01.20,10.80) {{\scriptsize TX turnaround}};

% Preamble  02.00 -- 04.00
\draw[fill=blue!40] (02.00,10.00) rectangle (04.00,10.50);
\draw[dashed] (02.50,10.00) -- (02.50,10.50);
\draw[dashed] (03.00,10.00) -- (03.00,10.50);
\draw[dashed] (03.50,10.00) -- (03.50,10.50);
\draw[*-] (02.75,10.25) -- (02.60,11.25);
\node[left] at (03.50,11.40) {{\scriptsize preamble}};

% SFD 04.00 -- 04.50
\draw[fill=green] (04.00,10.00) rectangle (04.50,10.50);
\draw[*-] (04.25,10.25) -- (04.00,10.90);
\node[left] at (04.10,11.00) {{\scriptsize SFD}};

% T1 04.50
\draw[red,ultra thick,->] (04.50,11.50) -- (04.50,10.50);
\node[right] at (04.20,11.80)
    {\textcolor{red}{$\tau_{i,\texttt{start}}$}};

% Payload 04.50 --  07.00
\draw[fill=orange!80] (04.50,10.00) rectangle (05.00,10.50);
\draw[*-] (04.75,10.25) -- (05.10,11.25);
\node[right] at (05.00,11.40) {{\scriptsize length}};
\draw[fill=yellow!60] (05.00,10.00) rectangle (06.00,10.50);
\draw[dashed] (05.50,10.00) -- (05.50,10.50);
\draw[*-] (05.75,10.25) -- (06.00,10.80);
\node[right] at (05.90,11.00) {{\scriptsize payload}};
\draw[fill=magenta!40] (06.00,10.00) rectangle (07.00,10.50);
\draw[dashed] (06.50,10.00) -- (06.50,10.50);
\draw[*-] (06.75,10.25) -- (07.50,10.40);
\node[right] at (07.40,10.50) {{\scriptsize CRC}};

% SFD2SFDlag
\draw[fill=black] (12.00,10.00) rectangle (12.30,10.30);

% T2 12.30
\draw[red,ultra thick,->] (12.30,10.30) -- (12.30,11.50);
\node[right] at (11.90,11.80)
    {\textcolor{red}{$\tau_{i,\texttt{end}}$}};

% time axis
\draw[thick,->] (-01.60,10.00) -- (21.00,10.00);
\node[right] at (-01.75,09.80) {\scriptsize \bf Initiator $i$};
\node[left] at (21.00,09.70) {\small time};

% CONNECTIONS ---------------------------------------------------------------------------------------------------

% Start of operation
\draw[ultra thick,dotted] (01.00,10.00) -- (01.00,07.00);
\node[right] at (01.00,07.65) {\scriptsize Start of};
\node[right] at (01.00,07.20) {\scriptsize operation};

% Packet to master
\draw (02.00,10.00) -- (02.50,08.00);
\draw (05.30,08.00) -- (05.30,09.00);
\draw[red,ultra thick,->] (04.50,10.00) -- (05.00,08.00);
\draw (07.00,10.00) -- (07.50,08.00);
\draw (17.50,08.00) -- (18.00,10.00);

% Master delay
\draw (09.00,08.00) -- (09.00,09.00);
\draw[red,ultra thick,->] (05.30,08.80) -- (09.00,08.80);
\node[right] at (05.30,09.10)
    {\textcolor{red}{\scriptsize Retransm. delay $\tau_w$}};

% Paclet to slave
\draw (09.00,08.00) -- (09.50,10.00);
\draw[red,ultra thick,->] (11.50,08.00) -- (12.00,10.00);

% End of operation
\draw[ultra thick,dotted] (18.00,08.00) -- (18.00,11.00);
\node[right] at (18.00,10.80) {\scriptsize End of};
\node[right] at (18.00,10.35) {\scriptsize operation};

% MASTER --------------------------------------------------------------------------------------------------------------

% SFD2SFDlag
\draw[fill=black] (05.00,08.00) rectangle (05.30,07.70);

% TX turnaround  08.00 -- 09.00
\draw[fill=gray!20] (08.00,08.00) rectangle (09.00,07.70);
\draw[*-] (08.25,07.92) -- (08.00,07.35);
\node[left] at (08.10,07.20) {{\scriptsize TX turnaround}};

% Preamble  09.00 -- 11.00
\draw[fill=blue!40] (09.00,08.00) rectangle (11.00,07.50);
\draw[dashed] (09.50,08.00) -- (09.50,07.50);
\draw[dashed] (10.00,08.00) -- (10.00,07.50);
\draw[dashed] (10.50,08.00) -- (10.50,07.50);
\draw[*-] (09.75,07.75) -- (09.20,06.80);
\node[left] at (09.30,06.70) {{\scriptsize preamble}};

% SFD  11.00 -- 11.50
\draw[fill=green] (11.00,08.00) rectangle (11.50,07.50);
\draw[*-] (11.25,07.75) -- (11.00,06.80);
\node[left] at (11.10,06.70) {{\scriptsize SFD}};

% Payload 11.50 --  12.00
\draw[fill=orange!80] (11.50,08.00) rectangle (12.00,07.50);
\draw[*-] (11.75,07.75) -- (12.10,06.80);
\node[right] at (12.05,06.70) {{\scriptsize length}};

\draw[white,fill=yellow!60] (12.00,08.00) rectangle (13.00,07.50);
\draw (12.00,08.00) -- (12.00,07.50) -- (13.00,07.50);
\draw[dashed] (12.50,08.00) -- (12.50,07.50);
\node[right] at (12.95,07.70) {{\scriptsize payload (bar graph)}};

\draw[white,fill=yellow!60] (16.50,08.00) rectangle (17.50,07.50);
\draw (16.50,07.50) -- (17.50,07.50) -- (17.50,08.00);
\draw[dashed] (17.00,08.00) -- (17.00,07.50);

% time axis
\draw[thick,->] (-01.60,08.00) -- (21.00,08.00);
\node[right] at (-01.75,07.80) {\scriptsize \bf Predecessor $p$};
\node[left] at (21.00,07.70) {\small time};

% Meas and result

\draw[|-|, thick, red] (4.50,14.20) -- (12.30,14.20);
\node[right] at (12.90,14.20)
    {\textcolor{red}{\small Measured time interval}};
\draw[dashed,red] (4.50,14.20) -- (4.50,12.20);
\draw[dashed,red] (12.30,14.20) -- (12.30,12.20);

\draw[|-|, thick, blue] (05.30,13.60) -- (09.00,13.60);
\draw[|-|, thick, blue] (05.30,13.60) -- (11.50,13.60);
\node[right] at (12.90,13.60)
    {\textcolor{blue}{\small Known time intervals}};

\draw[|-|, thick] (05.00,13.00) -- (05.30,13.00);
\draw[|-|, thick] (12.00,13.00) -- (12.30,13.00);
\node[right] at (12.90,13.00)
    {\small SFD detection lag};

\draw[|-|, thick, green!60!black] (04.50,12.40) -- (05.00,12.40);
\draw[|-|, thick, green!60!black] (11.50,12.40) -- (12.00,12.40);

\node[right, green!60!black] at (12.90,12.40)
    {\small 2 $\times$ propagation delay};

\end{tikzpicture}
 \end{center}
 \caption{Overall timeline for the proposed delay measurement scheme.}
 \label{fig:Timeline}
\end{figure*}

It is possible to reserve a short time interval after each flooding,
to be used for propagation delay estimation. During this time
interval, the MAC protocol used by the nodes in ordinary operations
needs to be disabled as done during flooding. The interval is composed
of $s$ time slots, with $1 \leq s \leq n$. During these time slots,
using a TDMA scheme, each node can estimate its propagation delay from
the nodes in the previous hop, as explained in
Section~\ref{sec:RoundTrip}. Since $n$ is known and each node knows
its id, a simple round-robin scheme can be used for the TDMA. This
means that all the nodes periodically estimate their propagation delay
from the previous hop every $\lceil n / s \rceil$ synchronization
periods. The parameter $s$ allows the scheme to trade off the radio bandwidth
usage (and thus power consumption) for the speed at which a node
becomes aware of propagation delay changes and, as will be shown in the following, the time required for network formation.

The overall operation of the scheme for one of the TDMA slots is
summarised in Figure~\ref{fig:Timeline}. During each of the $k$ time slots a single node (hereinafter, $i$, as the \emph{initiator}), belonging to hop
$h$, can send a propagation delay request packet.
This packet is sent in broadcast, and the node timestamps its local Start Frame Delimiter (SFD) occurring at $\tau_{i,\texttt{start}}$.
The request has a two-byte payload. The first byte is a packet type field
identifying it as a round-trip request, and the second byte is $h-1$.
The nodes whose hop number is not $h-1$ ignore the packet.

The nodes in the predecessor
set $\mathcal{P}_i$ act collectively as the \emph{predecessor} $p$,
thanks to constructive interference. The predecessor, upon getting the
request, waits for a fixed retransmission delay $\tau_w$, known
network-wide and used to account for the transceiver turnaround and the duration of the request packet, as well as to absorb any software-induced jitter. The predecessor
sends its response packet, and the initiator $i$ timestamps the
corresponding SFD (occurring at $\tau_{i,\texttt{end}}$). The response
packet contains the predecessor's cumulated delay $\delta_{M
  \rightarrow p}$, encoded in bar graph form.
Note that there is no need to include a timestamp neither in
the request nor in the response packets.

Node $i$ then takes the difference
$\tau_{i,\texttt{end}}-\tau_{i,\texttt{start}}$, and subtracts the
retransmission delay $\tau_w$, the duration of a four-byte preamble
plus SFD, and an additional short time -- a transceiver-specific
parameter that can be easily measured in a laboratory setting -- for
the lag in the SFD detection. The result, as evidenced in
Figure~\ref{fig:Timeline}, is twice the propagation delay from the
predecessor, whence the measurement of $\delta_{p \rightarrow i}$.
Finally, by inspecting the response packet content, the initiator obtains the value of $\delta_{M \rightarrow p}$, therefore completing the estimate of the full delay from the master $\delta_{M \rightarrow i}$.

This process is prone to three main source of errors:
\begin{itemize}
\item possible variations in the propagation delay due for example to
  scattering caused by moving obstacles;
\item jitter in the SFD lag entity;
\item quantization in turning the round-trip time in a measurement counted in
  clock ticks.
\end{itemize}

The first error source is highly environment-dependent, and hardly any
general consideration can be made on it. That is why in the following
we show both indoor and outdoor experiments, testifying that the
caused errors are within a tolerable range.
For the latter two causes, the SFD detection
lag has a small variance~\cite{bib:SchmidEtAl-2010b}, so the resulting
error is comparable with one tick of the counting clock.
Since we used off-the-shelf components, we can conjecture that our
finding is general.

Using the estimate of the cumulated delay $\delta_{M \rightarrow i}$
node $i$ can compute a compensation term $c_i(k)$ at time instant $k$
($k$ counts the number of times the node has transmitted a propagation
delay request packet).
The compensation term $c_i(k)$ is then applied
at the clock of node $i$ to enhance synchronization.

First, $c_i(k)$ is subjected to a sanity check to eliminate evident
outliers, verifying for example that the round-trip delay from the
last hop is not outside the radio range. Despite this sanity check,
two other issues should be taken into account. First, it is important
to reduce as much as possible the jitter of the compensation term caused
by measurement errors and moving obstacles.
Second, synchronization schemes like FLOPSYNC-2~\cite{bib:TerraneoEtAl-2014a}
guarantee clock monotonicity.
In the application of the correction term $c_i(k)$ we should ensure to
preserve this property.

The first issue is solved using a lowpass filter on the compensation
term. Denoting by $c_i(k)$ the compensation term computed by the
procedure of Figure~\ref{fig:Timeline}, the actually applied
$c^{\texttt{applied}}_i(k)$ is defined as
\begin{equation}
  c^{\texttt{applied}}_i(k) = a \cdot c^{\texttt{applied}}_i(k-1) + (1-a) \cdot c_i(k)
 \label{eqn:LPF-ck}
\end{equation}
when $k>0$. The first term $c^{\texttt{applied}}_i(0)$ is set to
$c_i(0)$ to speed up convergence with the available information, i.e, at node boot the first propagation delay measure is used as first guess for the filter initialization.
The unity minus the value of the filter pole $a \in [0,1)$ is interpreted as the
one-step attenuation for a pulse outlier: For example, setting
$a=0.75$ causes such an outlier to be attenuated by a factor of four.
Notice that the cumulated propagation delay $\delta_{M \rightarrow p}$ that is sent to the next hop is the filtered value, to counteract the accumulation of jitter across multiple hops.

For the second issue, the FLOPSYNC-2 virtual clock is corrected by a
quantity that starts from zero at the instant of the generic $k$-th
measurement, and reaches $c_i(k)$ exponentially within one FLOPSYNC-2
synchronization period $T$. This is obtained with an additional first-order
lowpass filter in the same form as~\eqref{eqn:LPF-ck}, its pole being
computed so that the slope of the virtual clock never goes below a
given percentage of the slope forecast by FLOPSYNC-2.

\subsection{Network formation}

So far we assumed the network to be formed. The last remark for the
full scheme is about network formation. When a node first boots, it
waits its turn in the TDMA schedule, and then sends a propagation
delay request packet. If it receives an answer from nodes in its
predecessor set, it initializes the filter and starts answering to
propagation delay requests for the next hop. In case a node receives
a request but has not yet received its cumulated delay from at least
one node of its predecessor set, the node does not answer the request.
When this happens, within the first $\lceil n / s \rceil$
synchronization periods all nodes of the first hop had a chance to
measure their delay from the master, and thus will be able to respond
to future requests from the next hop nodes. In the worst case, the time
it takes for all the nodes of a network of $h_{\max}$ hops to become
aware of propagation delays is $\lceil n / s \rceil \cdot h_{\max}$
synchronization periods. 

\section{Experimental results}
\label{sec:ExpResults}

This section shows experimental results to support our proposal. We
run our delay compensator on a \ac{WSN} composed of nodes built around
a CC2520 transceiver and an ARM Cortex-M3 microcontroller running at
24$\,$MHz, with a timestamping resolution of 42$\,$ns. The software is
written in C++, as an application for the Miosix operating
system\footnote{\url{http://miosix.org/}} and available for
download\footnote{\url{http://miosix.org/flopsync.html}}. In the experimental
assessment we try to test for the worst conditions, in some cases also
forcing the network to behave in a worse way than its normal behavior
(for example preventing nodes to exploit some potential edges or testing constructive interference with skew and interferences beyond reasonable values).

\subsection{Measuring the single-hop propagation delay}

The first set of experiments assesses the viability of measuring
propagation delays using round-trip measurements with off-the-shelf
microcontrollers and radio transceivers. The task is challenging as
individual nodes in \acp{WSN} are often placed at small distances --
although a multi-hop network can be quite large -- thus requiring
high resolution time measurements to estimate the previous hop delay.

We used the microcontroller available resources, implementing
hardware-timed packet transmission and hardware-based packet arrival
timestamping, as done for FLOPSYNC-2~\cite{bib:TerraneoEtAl-2014a}.
This allows us to control the radio with a time granularity of one
timer tick. The inevitable noise in the measurements, in the form of
time jitter, turns here to our advantage, as it permits to sample
the propagation delay below the quantization limit imposed by the
timestamping resolution~\cite{bib:McDonnell-2011}. Thanks to the filtering
applied to the raw measures, as described in Section~\ref{sec:FullScheme}, the average
propagation delay measurement error was reduced below one timer tick ($42\,$ns).

\begin{figure}[t]
 \begin{center}
 \begin{tikzpicture}
\pgfplotsset{width=\columnwidth,height=3.5cm,every axis/.append style={font=\scriptsize}}
\begin{axis}[
/pgf/number format/.cd,
1000 sep={},
scaled y ticks = false,
xmin=-5,
xmax=75,
ymin=-0.00000005,
ymax= 0.00000005,
xticklabels={0m,10m,20m,30m,40m,50m,60m,70m},
xtick={0,10,20,30,40,50,60,70},
ytick={-0.00000004,-0.00000002,0,0.00000002,0.00000004},
yticklabels={-40ns,-20ns,0,+20ns,+40ns},
xlabel={Nodes Distance},
xlabel style={yshift=0.4em},
ylabel={},
ylabel style={yshift=-0.8em},
xlabel near ticks,
ylabel near ticks,
]
\addplot[
		red,
        error bars/.cd,
        y dir=both,
        y explicit,
    ] table[x index=0,y index=1,y error index=2,col sep=comma] {img/tikzpgfplots/IndoorPropagationDelay.csv}; 

\addplot+[black,mark=*,mark options={fill=black}] table[x index=0,y index=1,y error index=2,col sep=comma] {img/tikzpgfplots/IndoorPropagationDelay.csv};
% \node[above,yshift=1mm,fill=white,inner sep=2pt] at (axis cs:1,0) {\scriptsize   50ns};
% \node[above,xshift=-1mm,yshift=-6mm,rotate=90,inner sep=2pt] at (axis cs:1,0) {\scriptsize  866ns};
% 
% \node[above,yshift=1mm,fill=white,inner sep=2pt] at (axis cs:2,0) {\scriptsize   26ns};
% \node[above,xshift=-1mm,yshift=-6mm,rotate=90,inner sep=2pt] at (axis cs:2,0) {\scriptsize  858ns};
% 
% \node[above,yshift=1mm,fill=white,inner sep=2pt] at (axis cs:3,0) {\scriptsize  -25ns};
% \node[above,xshift=-1mm,yshift=-6mm,rotate=90,inner sep=2pt] at (axis cs:3,0) {\scriptsize  858ns};
% 
% \node[above,yshift=1mm,fill=white,inner sep=2pt] at (axis cs:4,0) {\scriptsize  -12ns};
% \node[above,xshift=-1mm,yshift=-6mm,rotate=90,inner sep=2pt] at (axis cs:4,0) {\scriptsize  877ns};
% 
% \node[above,yshift=1mm,fill=white,inner sep=2pt] at (axis cs:5,0) {\scriptsize   29ns};
% \node[above,xshift=-1mm,yshift=-6mm,rotate=90,inner sep=2pt] at (axis cs:5,0) {\scriptsize  873ns};
% 
% \node[above,yshift=1mm,fill=white,inner sep=2pt] at (axis cs:6,0) {\scriptsize  -62ns};
% \node[above,xshift=-1mm,yshift=-6mm,rotate=90,inner sep=2pt] at (axis cs:6,0) {\scriptsize  880ns};
% 
% \node[above,yshift=1mm,fill=white,inner sep=2pt] at (axis cs:7,0) {\scriptsize -113ns};
% \node[above,xshift=-1mm,yshift=-6mm,rotate=90,inner sep=2pt] at (axis cs:7,0) {\scriptsize  891ns};
% 
% \node[above,yshift=1mm,fill=white,inner sep=2pt] at (axis cs:8,0) {\scriptsize -104ns};
% \node[above,xshift=-1mm,yshift=-6mm,rotate=90,inner sep=2pt] at (axis cs:8,0) {\scriptsize  888ns};

\addplot[densely dotted] coordinates {(-5,0) (75,0)};
\end{axis}
\end{tikzpicture}
 \caption{Propagation delay measurement error (indoor).}
 \label{fig:IndoorPropagationDelay}
 \end{center}
\end{figure}
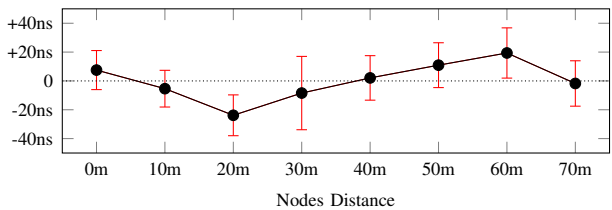
\begin{figure}[t]
 \begin{center}
 \begin{tikzpicture}
\pgfplotsset{width=\columnwidth,height=3.5cm,every axis/.append style={font=\scriptsize}}
\begin{axis}[
/pgf/number format/.cd,
1000 sep={},
scaled y ticks = false,
xmin=-10,
xmax=130,
ymin=-0.00000005,
ymax= 0.00000009,
xticklabels={0m,20m,40m,60m,80m,100m,120m},
xtick={0,20,40,60,80,100,120},
ytick={-0.00000004,0,0.00000004,0.00000008},
yticklabels={-40ns,0,+40ns,+80ns},
xlabel={Nodes Distance},
xlabel style={yshift=0.4em},
ylabel={},
ylabel style={yshift=-0.8em},
xlabel near ticks,
ylabel near ticks,
]
\addplot[
		red,
        error bars/.cd,
        y dir=both,
        y explicit,
    ] table[x index=0,y index=1,y error index=2,col sep=comma] {img/tikzpgfplots/OutdoorWithPeoplePropagationDelay.csv}; 

\addplot+[black,mark=*,mark options={fill=black}] table[x index=0,y index=1,y error index=2,col sep=comma] {img/tikzpgfplots/OutdoorWithPeoplePropagationDelay.csv};
% \node[above,yshift=1mm,fill=white,inner sep=2pt] at (axis cs:1,0) {\scriptsize   50ns};
% \node[above,xshift=-1mm,yshift=-6mm,rotate=90,inner sep=2pt] at (axis cs:1,0) {\scriptsize  866ns};
% 
% \node[above,yshift=1mm,fill=white,inner sep=2pt] at (axis cs:2,0) {\scriptsize   26ns};
% \node[above,xshift=-1mm,yshift=-6mm,rotate=90,inner sep=2pt] at (axis cs:2,0) {\scriptsize  858ns};
% 
% \node[above,yshift=1mm,fill=white,inner sep=2pt] at (axis cs:3,0) {\scriptsize  -25ns};
% \node[above,xshift=-1mm,yshift=-6mm,rotate=90,inner sep=2pt] at (axis cs:3,0) {\scriptsize  858ns};
% 
% \node[above,yshift=1mm,fill=white,inner sep=2pt] at (axis cs:4,0) {\scriptsize  -12ns};
% \node[above,xshift=-1mm,yshift=-6mm,rotate=90,inner sep=2pt] at (axis cs:4,0) {\scriptsize  877ns};
% 
% \node[above,yshift=1mm,fill=white,inner sep=2pt] at (axis cs:5,0) {\scriptsize   29ns};
% \node[above,xshift=-1mm,yshift=-6mm,rotate=90,inner sep=2pt] at (axis cs:5,0) {\scriptsize  873ns};
% 
% \node[above,yshift=1mm,fill=white,inner sep=2pt] at (axis cs:6,0) {\scriptsize  -62ns};
% \node[above,xshift=-1mm,yshift=-6mm,rotate=90,inner sep=2pt] at (axis cs:6,0) {\scriptsize  880ns};
% 
% \node[above,yshift=1mm,fill=white,inner sep=2pt] at (axis cs:7,0) {\scriptsize -113ns};
% \node[above,xshift=-1mm,yshift=-6mm,rotate=90,inner sep=2pt] at (axis cs:7,0) {\scriptsize  891ns};
% 
% \node[above,yshift=1mm,fill=white,inner sep=2pt] at (axis cs:8,0) {\scriptsize -104ns};
% \node[above,xshift=-1mm,yshift=-6mm,rotate=90,inner sep=2pt] at (axis cs:8,0) {\scriptsize  888ns};

\addplot[densely dotted] coordinates {(-10,0) (130,0)};
\end{axis}
\end{tikzpicture}
 \end{center}
 \caption{Propagation delay measurement error (outdoor).}
 \label{fig:OutdoorWithPeoplePropagationDelay}
\end{figure}
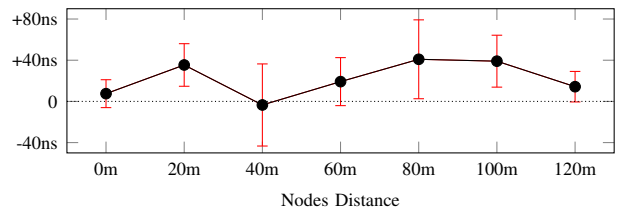

Our experimental setup consists in two nodes, serving as initiator
and predecessor, exchanging packets as shown in
Figure~\ref{fig:Timeline}. For all the experiments, we positioned the
nodes and left them in the same position for five minutes, collecting
propagation delay measurements every second. We then computed the
error as the measured value minus the nominal one, computed dividing
the (known) distance by the speed of radio waves in air.
We repeated the experiment in two different conditions. The first 
experiment is performed indoor, in a corridor, varying the distances
of the nodes by $10\,$m for each experiment. In indoor conditions, a few
people interacted, randomly, with the setup.
Figure~\ref{fig:IndoorPropagationDelay} shows the mean value and
standard deviation of the computed error for different distances.
The second experiment is performed outdoor, on a street pavement,
varying the nodes distances with a $20\,$m step. In this
second case, many people and vehicles were moving around during the
experiment, thereby making scattering relevant.
Figure~\ref{fig:OutdoorWithPeoplePropagationDelay} summarizes the
results.

As can be seen, the indoor experiment results in average errors
below one timer tick, while in the outdoor experiment, the scheme
tends to overestimate the nominal propagation delay. This is not
surprising, as the presence of people between
nodes obstructed their line of sight, forcing radio waves to
follow a longer path. In both cases the technique allows for delay
measurement at the clock tick timescale in a reliable manner. To the best of the authors' knowledge, this was not possible to date with off-the-shelf hardware.

\subsection{Measuring the multi-hop propagation delay}

We also performed multi-hop propagation delay experiments, to assess
how the delay accumulates as the number of hops increases. The
experimental setup resembles the previous one, but uses multiple nodes
and performs measurements using the TDMA schedule described in
Section~\ref{sec:FullScheme}. Every node, except the master, acts
alternatively as initiator to measure the propagation delay from the
previous hop, and as predecessor for the next hop. Although bar graph
encoding was used to transfer the cumulated propagation delay, a
single node per hop was used. The efficacy of the bar graph
encoding is tested separately in the following.

\begin{figure}[t]
 \begin{center}
  \begin{tikzpicture}
\pgfplotsset{width=\columnwidth,height=3.5cm,every axis/.append style={font=\scriptsize}}
\begin{axis}[
/pgf/number format/.cd,
1000 sep={},
scaled y ticks = false,
xmin=0.5,
xmax=6.5,
ymin=-0.00000011,
ymax= 0.00000006,
xticklabels={1,2,3,4,5,6},
xtick={1,2,3,4,5,6},
ytick={-0.00000010,-0.00000005,0,0.00000005},
yticklabels={-100ns,-50ns,0,+50ns},
xlabel={hop},
xlabel style={yshift=0.4em},
ylabel={},
ylabel style={yshift=-0.8em},
xlabel near ticks,
ylabel near ticks,
]
\addplot[
		red,
        error bars/.cd,
        y dir=both,
        y explicit,
    ] table[x index=0,y index=1,y error index=2,col sep=comma] {img/tikzpgfplots/MultiHopPropagationDelay.csv}; 

\addplot+[black,mark=*,mark options={fill=black}] table[x index=0,y index=1,y error index=2,col sep=comma] {img/tikzpgfplots/MultiHopPropagationDelay.csv};
% \node[above,yshift=1mm,fill=white,inner sep=2pt] at (axis cs:1,0) {\scriptsize   50ns};
% \node[above,xshift=-1mm,yshift=-6mm,rotate=90,inner sep=2pt] at (axis cs:1,0) {\scriptsize  866ns};
% 
% \node[above,yshift=1mm,fill=white,inner sep=2pt] at (axis cs:2,0) {\scriptsize   26ns};
% \node[above,xshift=-1mm,yshift=-6mm,rotate=90,inner sep=2pt] at (axis cs:2,0) {\scriptsize  858ns};
% 
% \node[above,yshift=1mm,fill=white,inner sep=2pt] at (axis cs:3,0) {\scriptsize  -25ns};
% \node[above,xshift=-1mm,yshift=-6mm,rotate=90,inner sep=2pt] at (axis cs:3,0) {\scriptsize  858ns};
% 
% \node[above,yshift=1mm,fill=white,inner sep=2pt] at (axis cs:4,0) {\scriptsize  -12ns};
% \node[above,xshift=-1mm,yshift=-6mm,rotate=90,inner sep=2pt] at (axis cs:4,0) {\scriptsize  877ns};
% 
% \node[above,yshift=1mm,fill=white,inner sep=2pt] at (axis cs:5,0) {\scriptsize   29ns};
% \node[above,xshift=-1mm,yshift=-6mm,rotate=90,inner sep=2pt] at (axis cs:5,0) {\scriptsize  873ns};
% 
% \node[above,yshift=1mm,fill=white,inner sep=2pt] at (axis cs:6,0) {\scriptsize  -62ns};
% \node[above,xshift=-1mm,yshift=-6mm,rotate=90,inner sep=2pt] at (axis cs:6,0) {\scriptsize  880ns};
% 
% \node[above,yshift=1mm,fill=white,inner sep=2pt] at (axis cs:7,0) {\scriptsize -113ns};
% \node[above,xshift=-1mm,yshift=-6mm,rotate=90,inner sep=2pt] at (axis cs:7,0) {\scriptsize  891ns};
% 
% \node[above,yshift=1mm,fill=white,inner sep=2pt] at (axis cs:8,0) {\scriptsize -104ns};
% \node[above,xshift=-1mm,yshift=-6mm,rotate=90,inner sep=2pt] at (axis cs:8,0) {\scriptsize  888ns};

\addplot[densely dotted] coordinates {(0.5,0) (6.5,0)};
\end{axis}
\end{tikzpicture}
 \end{center}
 \caption{Multi-hop propagation delay measurement error.}
 \label{fig:MultiHopPropagationDelay}
\end{figure}
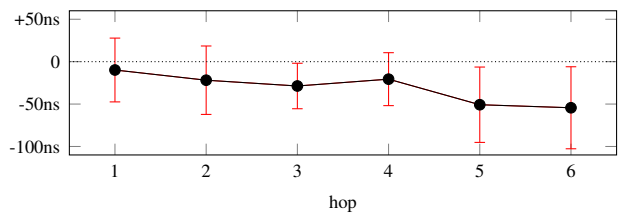

Figure~\ref{fig:MultiHopPropagationDelay} shows the propagation delay
measurement error and standard deviation as a function of the hop
count. In this experiment all nodes were placed at a multiple of
$68\,$m --- e.g., the sixth hop is $408\,$m away from the master.
Although the propagation delay measurement error increases with the
hop count, the \emph{relative} error, i.e, the average error divided
by the total propagation delay remains fairly constant and always
below 5\% of the real value. Our proposal therefore cancels at least
95\% of the error induced by propagation delays for a clock
synchronization scheme. Moreover, the standard deviation does not
increase with the hop count.

\subsection{Combining responses with the bar graph encoding}

This section shows the capabilities of the proposed bar-graph
encoding. We tried to maximize the repeatability of these experiments. 
For that, we engineered a special node by connecting a microcontroller
to three transceivers. Two of these transceivers transmitted in bar
graph form similar but not equal numbers with the same transmission
power. Packets were transmitted skewed by a variable time $\Delta t$.
The third transceiver was configured to concurrently transmit with
less power (to emulate a longer distance) a different number, thus
acting as an interferer. This setting is representative of the
situation described in Figure~\ref{fig:RoundTrip-1}, with a
predecessor set of two near nodes and a set of three responding nodes,
the third one being shadowed by distance.

\begin{figure}[t]
 \begin{center}
  \includegraphics[width=0.95\columnwidth]{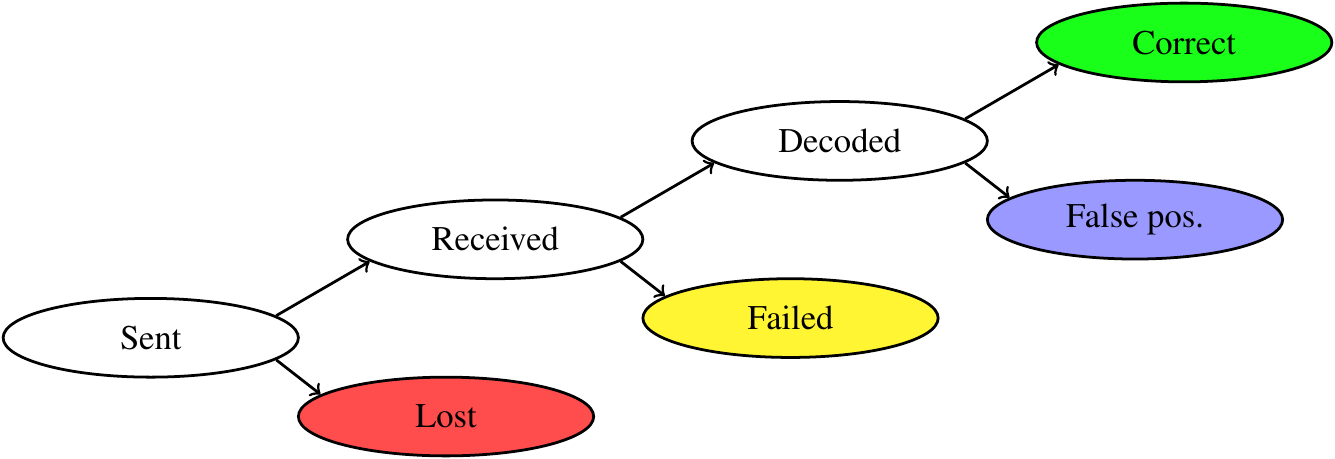}
 \end{center}
 \caption{Classification of packets for the bar graph test.}
 \label{fig:BarGraphTest-PacketTree}
\end{figure}

We then used two further nodes to receive data from the predecessor
set. The first one was located at a distance of $5\,$m with respect to
the transmitting setup, the other at $60\,$m. The choice of both a
near and a far predecessor further generalizes the results.

The test involved sending roughly $6 \cdot 10^5$ packets, divided in
all the possible combination of transmission skew between the
equally powerful transmitters, in the set $\{80\,$ns, $160\,$ns, 
$320\,$ns, $640\,$ns$\}$, and interferer power in the set $\{$off,
$-18\,$dBm, $-7\,$dBm, $-2\,$dBm$\}$. As illustrated by the decision
tree represented in Figure~\ref{fig:BarGraphTest-PacketTree}, the
outcomes of said transmissions are classified in four categories:
\begin{itemize}[noitemsep,topsep=0pt]
\item \emph{correct}, decoded and yielding a response within the
  values transmitted by the predecessor set nodes;
\item \emph{false positive}, causing the decoding algorithm to
  succeed but to output a number not in the expected range (most
  frequently, matching the interferer);
\item \emph{failed}, causing the algorithm to report the packet as too
  corrupted to be decoded;
\item \emph{lost}, not received for any reason.
\end{itemize}

\begin{figure}[h]
 \begin{center}
  \vspace{3mm}\hspace{-7mm}\includegraphics[width=1.07\columnwidth]{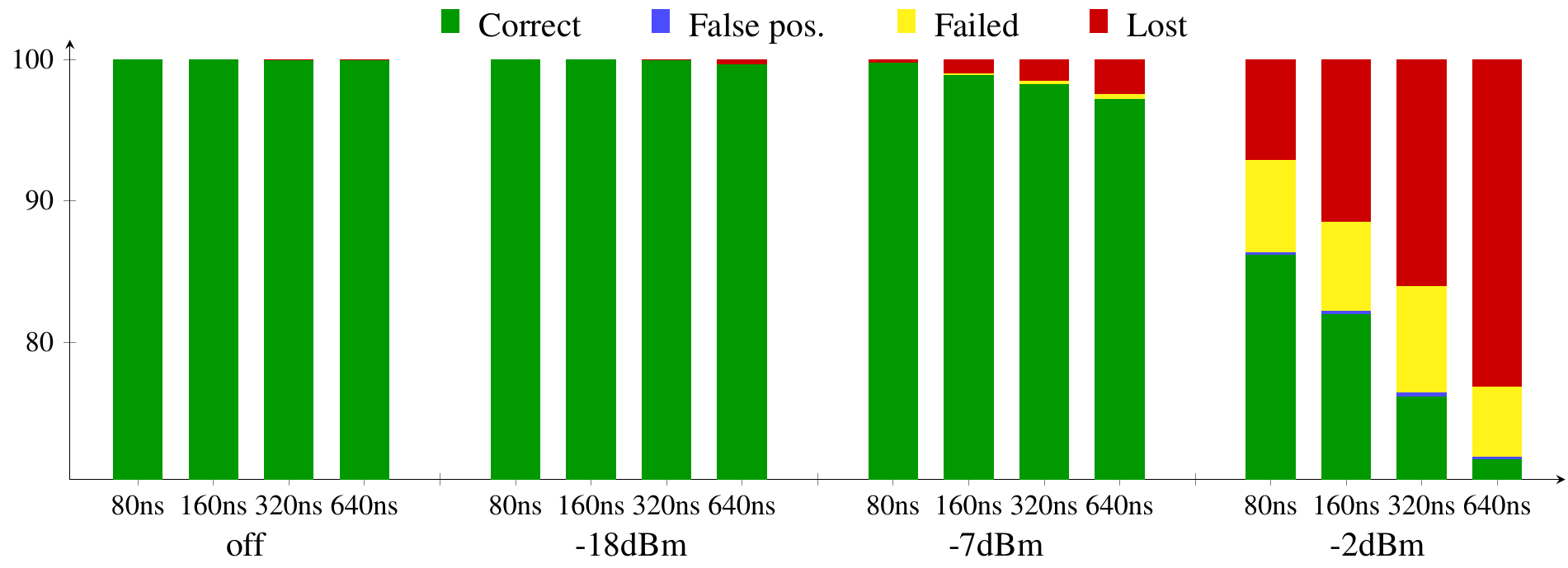}
  \vspace{3mm}\hspace{-7mm}\includegraphics[width=1.07\columnwidth]{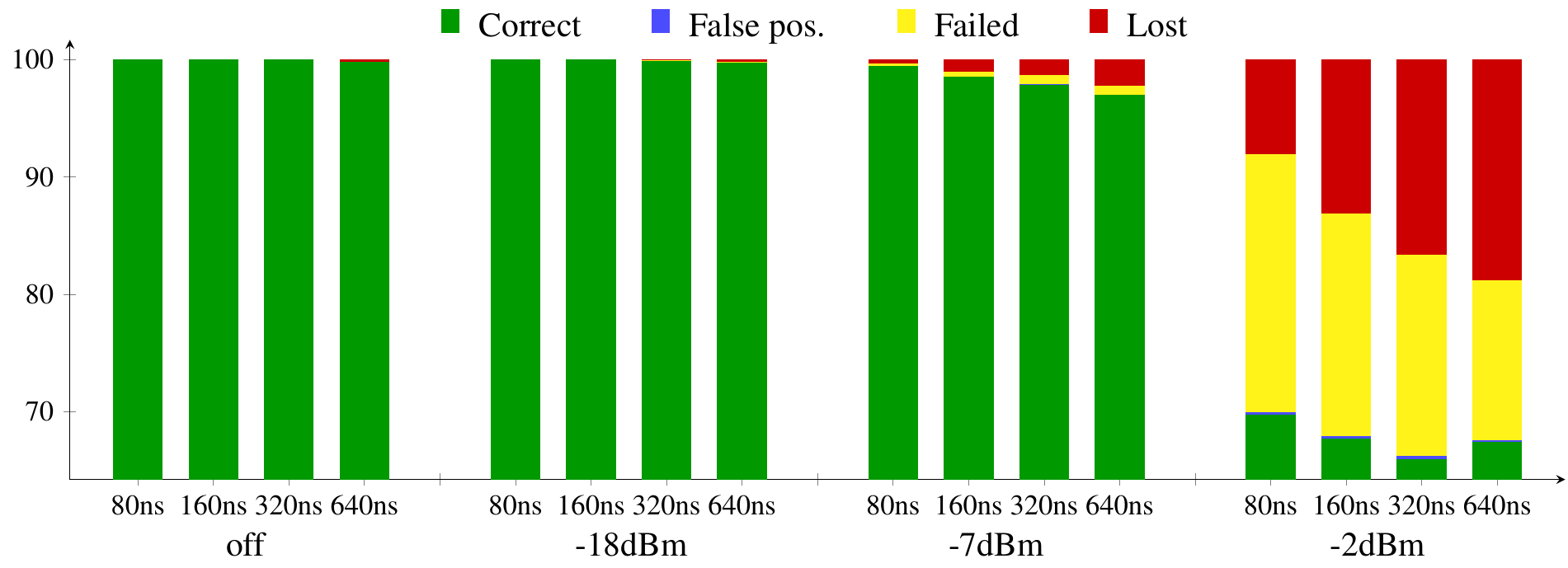}
 \end{center}
 \caption{Bar graph test -- receiver node placed at 5m from the transmitter, 16 bytes payload (top) and 64 bytes payload (bottom).}
 \label{fig:Interference-near}
\end{figure}

\begin{figure}[h]
 \begin{center}
  \vspace{3mm}\hspace{-7mm}\includegraphics[width=1.07\columnwidth]{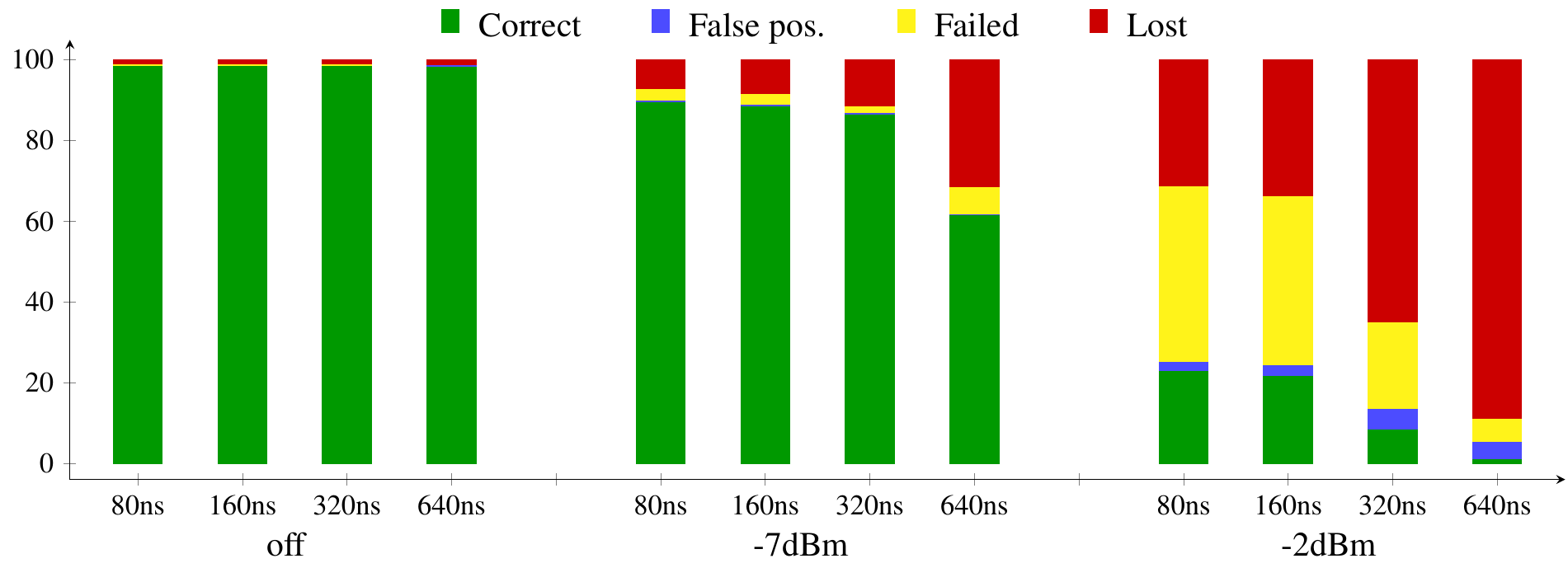}
  \vspace{3mm}\hspace{-7mm}\includegraphics[width=1.07\columnwidth]{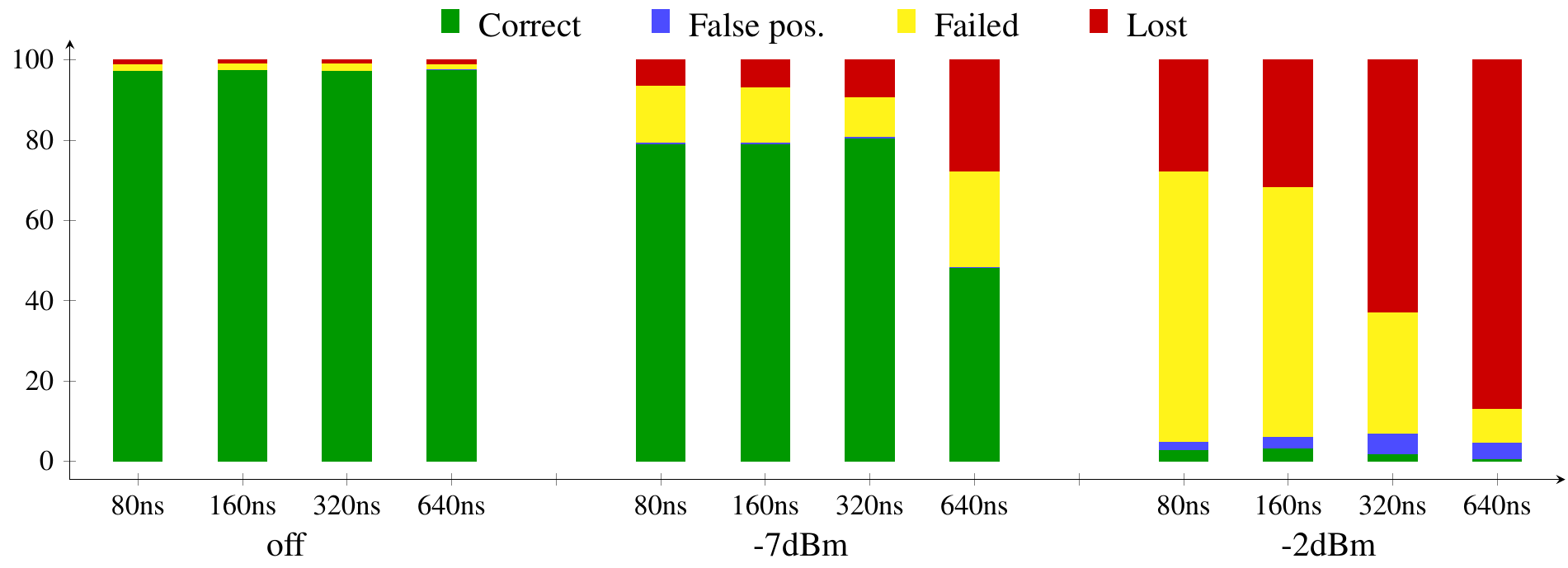}
 \end{center}
 \caption{Bar graph test -- receiver node placed at 60m from the transmitter, 16 bytes payload (top) and 64 bytes payload (bottom).}
 \label{fig:Interference-far}
\end{figure}

Results are reported as the percentage of packets falling into each
category. Figure~\ref{fig:Interference-near} shows the result for a
node distance of $5\,$m with a packet length of 16 and 64 Bytes, while
Figure~\ref{fig:Interference-far} (mind the different vertical scales) reports the results for a $60\,$m
distance.
In Figure~\ref{fig:Interference-far} the power of $-18\,$dBm is not reported as this would be equivalent to putting the interferer outside the radio range of the receiver, thus to the off case.

Considering the results for the $5\,$m distance, when the interfering
power is compatible with a significant distance difference (up to
$-7\,$dBm) and the transmission skew is within the Glossy tolerance of
$0.5\,\mu$s~\cite{bib:FerrariEtAl-2011a}, more than 97\% of the  packets are decoded correctly, regardless of payload
length. Results for the $60\,$m distance show a more relevant
difference between short and long packets, with the 16 Bytes case
resulting in at least 86\% correct packet reception, and the 64 Bytes
dropping to a minimum of 80\%.

Indeed, the technique fails only if either of the conditions above is
violated. However, a skew above $500\,\mu$s would undermine the
applicability of the Glossy flooding scheme itself, and can be easily
avoided with a well timed packet retransmission delay. For what
concerns a comparable power interferer, this is ruled out by
geometrical reasons. In detail, using a model for the attenuation of
radio waves in air~\cite{bib:Saunders-2007a}, a node to have a $-2\,$dB
difference from a node at $5\,$m from the receiver would need to be
located at $6.3\,$m, thus having a distance from the predecessor set
of only $1.3\,$m. In such a case, its propagation delay would not be
different from the other nodes in the predecessor set, and thus it
would not send a packet with a different number causing interference.
Applying the same reasoning to the case where nodes are spaced $60\,$m
apart results in the node sending at $-2\,$dBm being $15.5\,$m apart
from the predecessor set, a difference of less than two timer ticks.
Also in this case, the node would not cause interference. In practice,
therefore, we can rule out the failing conditions.

\subsection{Assessing the full scheme}

The last experiment considers the integration of the proposed
propagation delay compensator with a clock synchronization scheme, to
assess the achieved improvement. A total of seven nodes were used for
this test, using the FLOPSYNC-2 scheme configured with the default
parameters for indoor use, a the synchronization period of $60\,$s,
and $\alpha = 3/8$~\cite{bib:TerraneoEtAl-2014a}.

\begin{figure}[t]
 \begin{center}
  \vspace{3mm}\includegraphics[width=0.95\columnwidth]{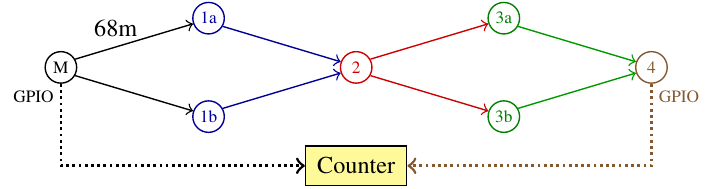}
 \end{center}
 \caption{Network topology for the full scheme assessment.}
 \label{fig:DoubleDiamond}
\end{figure}

The \emph{logical} network topology of the setup is described in
Figure~\ref{fig:DoubleDiamond}. This setup was chosen to have a
significant cumulated propagation delay from the master node to the
last hop. Moreover, two nodes were employed for Hop 1 and 3 to test
the effect of constructive interference in round-trip packet
exchanges. For this experiment the flooding scheme was slightly
altered, to force in software the network topology by manually assigning
each node to a hop. This allowed to fold the logical topology in order
to have the master node and node 4 physically next to each other,
while forcing the flooded packets to follow the entire $272\,$m path.
The application software running on the network periodically raised a
pin on the microcontroller (in a hardware timed way, avoiding software
jitter) at prescribed intervals. Having the master and last node
close together allowed to connect their pins to an SR620 frequency
counter, configured in time interval measurement mode to log the
clock synchronization error. This counter has a sub-nanosecond
resolution, far less than the measured time intervals.

\begin{table}[b]
\centering
\caption{Clock synchronization error at the fourth hop with and
  without propagation delay compensation.}
\label{tab:FullClockSync}
\begin{tabular}{lcc}
\hline \hline
                    & average                       & standard deviation \\
\hline \hline
plain FLOPSYNC-2    &                    1020$\,$ns & 696$\,$ns \\
enhanced FLOPSYNC-2 & \textcolor{white}{0}114$\,$ns & 687$\,$ns \\
\hline 
\end{tabular}
\end{table}

The experiment was repeated with the both plain FLOPSYNC-2, and with
FLOPSYNC-2 enhanced with the proposed propagation delay compensation
scheme. Table~\ref{tab:FullClockSync} shows how the lack of
propagation delay compensation causes the synchronization error of
plain FLOPSYNC-2 to exceed $1\,\mu$s, while the enhanced version
remains well into the sub-microsecond region. The standard deviation
does not increase significantly by propagation delay estimation.
Incidentally, the difference between the two averages multiplied by
the speed of radio waves in air amounts to $271.7\,$m, a value
remarkably close to the actual node distance.

\section{Related work}
\label{sec:RelWork}

Time synchronization protocols can be broadly categorized in two
classes: pairwise-synchronization schemes and flooding-based schemes.
TPSN~\cite{bib:GaneriwalEtAl-2004a} is one of the
most famous examples belonging to the first class.
TPSN needs to construct a spanning tree of the
network, and then performs synchronization along the edges. This
increases the overhead in terms of packet transmissions and thus
energy consumption, since packets should be sent for the spanning tree
creation and maintenance.
The availability of a spanning tree gives an explicit predecessor
information and allows in principle to estimate the propagation delay,
although this was not done as timestamping resolution was too limited
when the paper was published.
The second class, of which FTSP~\cite{bib:MarotiEtAl-2004a} is the
precursor, is based on broadcast messages, transmitted by a master
node to the neighboring ones and re-broadcast by the receivers for
nodes that are not in the master node radio range. Flooding-based
schemes are used because of their energy efficiency, since a single
transmission can synchronize multiple nodes simultaneously. Optimized
flooding schemes like Glossy~\cite{bib:FerrariEtAl-2011a} are crucial
for this approach. However, flooding based schemes have the
disadvantage that nodes do not know how the network is composed and
therefore cannot easily compensate for the propagation delay. 
Our proposal provides an efficient solution to this issue.

Some alternative techniques have been proposed to combine the best of
both worlds. Zeng et al.~\cite{Zeng2014ahn} proposed a measurement
architecture using distributed air sniffers, which provides convenient
transmission delay measurement, and requires no clock synchronization
or instrumentation at the node level. The problem of sniffer placement
still remains NP-hard~\cite{zeng2009milcom}, and the algorithms
proposed cannot be applied to large \acp{WSN}. Saifullah et
al.~\cite{saifullah2011rtas} analyze the effect of network delays on
WirelessHART networks. This is a specific case of \ac{WSN} for
industrial process monitoring and
control~\cite{abdelzaher2008performance, he2006rtas, Lu2002rts, Buttazzo2006RTR, lin2015tsn}. The proposed analysis methods are
however focused on obtaining upper bounds on the end-to-end delay of
every real-time periodic data flow in a \ac{WSN}. These contributions
further highlight the importance of efficiently estimating propagation
delays in the network.

The problem of estimating propagation delays in wireless
communication has been studied also in different areas, especially in
underwater acoustic sensor networks~\cite{guo2008wcmc, wen2013wcsp, huang2009tc}, and satellite communication~\cite{gun2009fcs, kito2012isap}.
The solutions proposed for these domains
are exploiting the entity of the delays -- exceeding the millisecond time
scale -- to improve the channel capacity. The problem that we faced in
this paper is conversely to compensate for sub-microsecond delays in
order to improve clock synchronization accuracy.

In this paper we have used interference to extract
information from colliding packets.
Collisions were also used to achieve indoor localization,
leveraging capture effect~\cite{vanvelzen2013percom}.
The idea of exploiting interference is similar to the one
presented in~\cite{dutta2008hotnets}, where the
initiator of a broadcast communication is able decode the
superposition of ACK packets thanks to the
constructive interference, although this case is simpler as
all interfering packets have the same payload.

Katti et al.~\cite{katti2007sigcomm} proposed a technique called
Analog Network Coding (ANC), that exploits signals transmission
instead of packet transmission, with a similar attitude to our
solution. Instead of trying to avoid interference in communication,
they exploit it to increase the channel capacity. ANC is based on the
idea that two senders can simultaneously send different packets on
the channel, allowing packets collision. Since in ANC signals are
transmitted, the collision of two signals results in a signal
corresponding to their sum.
The main limitation of this approach is that of requiring a
software-defined radio to gain access to the received signal and
disentangle the received packets. Our bar graph encoding, conversely,
works with commodity radio transceivers. In addition, ANC is limited
to the case of only two senders, while our proposal does not impose
restriction on the cardinality of the predecessor set.

\section{Conclusion}
\label{sec:conclusion}

In this paper we proposed a strategy to estimate the propagation delay
in a \ac{WSN} without the need to construct a spanning tree for the
network. Our strategy is based on a custom encoding of the cumulated
propagation delay that allows us to exploit the constructive
transmission interference also to send similar packets. Our estimation
strategy is here used to enhance flooding-based synchronization
schemes with a delay compensator. We implemented our strategy on top
of the FLOPSYNC-2 synchronization
scheme~\cite{bib:TerraneoEtAl-2014a}, achieving sub-microsecond clock
synchronization even in networks where propagation delays are
significant.

% \printbibliography

\end{document}